\documentclass[11pt,fleqn,twoside]{article}
\usepackage{amsfonts,amssymb,latexsym}
\makeatletter
\newcommand{\prava}[1]{\small\it
\begin{flushleft}
Copyright \copyright \ 1999 by  #1
\end{flushleft}}

\newcommand{\name}[1]{\begin{flushleft}
                       \LARGE \bf #1
                       \end{flushleft}\vspace{-3mm}}

\newcommand{\Author}[1]{\begin{flushleft}
                       \it #1 \end{flushleft}}

\newcommand{\Adress}[1]{\begin{flushleft}
                       \it #1 \end{flushleft}}

\newcommand{\Date}[1]{\begin{flushleft}
                      \small  \it #1 \end{flushleft}}

\newcommand{\ehkol}{Author \ name}
\newcommand{\ohkol}{Article \ name}
\renewcommand{\@evenhead}{
\hspace*{-3pt}\raisebox{-15pt}[\headheight][0pt]{\vbox{\hbox to \textwidth 
{\thepage \hfil \ehkol}\vskip4pt \hrule}}}
\renewcommand{\@oddhead}{
\hspace*{-3pt}\raisebox{-15pt}[\headheight][0pt]{\vbox{\hbox to \textwidth 
{\ohkol \hfil \thepage}\vskip4pt\hrule}}}
\renewcommand{\@evenfoot}{}
\renewcommand{\@oddfoot}{}

\newcommand{\be}{\begin{equation}}
\newcommand{\ee}{\end{equation}}
\newcommand{\ba}{\hspace*{-5pt}\begin{array}}
\newcommand{\ea}{\end{array}}

\newcommand{\ds}{\displaystyle}
\makeatother

\begin{document}

\thispagestyle{empty}
\setcounter{page}{332}
\renewcommand{\theequation}{\arabic{section}.\arabic{equation}}

\renewcommand{\ehkol}{F. Delduc and L. Gallot}
\renewcommand{\ohkol}{The Third Family of $N=2$ Supersymmetric 
KdV Hierarchies}

\begin{flushleft}
\footnotesize \sf
Journal of Nonlinear Mathematical Physics \qquad 1999, V.6, N~3,
\pageref{delduc-fp}--\pageref{delduc-lp}.
\hfill {\sc Article}
\end{flushleft}

\vspace{-5mm}

{\renewcommand{\footnoterule}{} 
{\renewcommand{\thefootnote}{} \footnote{\prava{F. Delduc and L. Gallot}}}

\name{A Note on the Third Family of {\mathversion{bold}$N=2$} Supersymmetric 
KdV Hierarchies}\label{delduc-fp}

\Author{F. DELDUC~$^\dag$ and L. GALLOT~$^\ddag$}
 
\renewcommand{\thefootnote}{$*$}

\Adress{$\dag$~Laboratoire de Physique\footnote{URA 1325 du CNRS, associ\'ee \`a 
l'Ecole Normale Sup\'erieure de Lyon.}, Groupe de Physique Th\'eorique ENS Lyon, \\
~~46 all\'ee d'Italie, 69364 Lyon CEDEX 07, France\\
~~E-mail: Francois.Delduc@ens-lyon.fr\\[1mm]
$\ddag$~Dipartimento di Fisica Teorica Universit\`a di Torino, \\
~~Via P. Giuria 1, 10125 Torino, Italy\\
~~E-mail: gallot@to.infn.it}

\Date{Received February 2, 1999; Revised February 24, 1999; Accepted  March 29, 1999}

\begin{abstract}
\noindent
We propose a hamiltonian formulation of the $N=2$
supersymmetric KP type hierarchy recently studied by Krivonos and Sorin.  
We obtain a quadratic hamiltonian structure which allows for several reductions of the KP 
type hierarchy. In particular, the third family of $N=2$ KdV hierarchies is recovered.
We also give an easy construction of Wronskian solutions of the KP and KdV type equations.
\end{abstract}

\section{Introduction}

The existence of three dif\/ferent $N=2$ supersymmetric integrable $n$-KdV hierarchies 
with the $N=2$ super ${\cal W}_n$ algebra as a hamiltonian structure has been made 
plausible by the works of many authors \cite{mathieu,pluri,yung2}. For two of these families, 
a complete description by means of a classical $r$-matrix approach using the algebra 
of chirality preserving pseudo-dif\/ferential operators ($\Psi$DOs) has been proposed 
in~\cite{DG1}. A formulation of the same hierarchies in the Drinfeld-Sokolov approach 
also exists~\cite{IK,DG2}.

The last remaining family of $N=2$ $n$-KdV hierarchies is of a somewhat dif\/ferent nature. 
Actually, the bosonic limit of the two f\/irst $N=2$ 
$n$-KdV hierarchies is composed of two decoupled KdV and non-standard KdV 
hierarchies~\cite{DG1,bokris} whereas the bosonic limit of the third one is the $(1,n)$ 
KdV hierarchy~\cite{bokris} which is irreducible~\cite{aratyn,BX,Dickey2}. Recently, 
Krivonos and Sorin~\cite{KS} gave a Lax representation for the third family of $N=2$
KdV hierarchies.

The aim of the present letter is to give the hamiltonian formulation of the $N=2$
KP type hierarchy which contains as reductions the above mentionned third family 
of $N=2$ $n$-KdV hierarchies and a third type of $N=2$ constrained KP systems as well. 
We shall proceed as follows. First, we define the KP type equations using the algebra  
${\cal D}$ of bosonic $\Psi$DOs with $N=2$ superf\/ields as their coef\/f\/icients. A peculiarity 
of these f\/lows on a $\Psi$DO~${\cal L}$ is that, although they are associated with a 
classical $r$-matrix on ${\cal D}$, they do not take the Lax form. Second, we show 
that, as was expected from \cite{KS}, the evolution equations of two $\Psi$DOs 
constructed from ${\cal L}$, one being chiral and the other antichiral, have the Lax form. 
This allows to determine an inf\/inite set of $N=2$ supersymmetric conserved quantities, 
or hamiltonians, for the KP type f\/lows. Third, we f\/ind a quadratic hamiltonian structure 
for the KP type hierarchy. Since the def\/ining f\/lows do not have the Lax form, this 
Poisson bracket does not have a standard form associated with a classical $r$-matrix. 
Hamiltonians are found to be in involution with respect to this Poisson bracket, 
which achieves the proof of integrability. Reductions to a f\/inite number of f\/ields 
and the bosonic limit are also brief\/ly discussed.

In the last part of this letter, we construct Wronskian solutions and $\tau$-functions 
of the KdV and KP type equations. This construction is a simple extension of the bosonic 
one. No such construction is known at present for the f\/irst two families of  $N=2$
KP and KdV hierarchies.}

\renewcommand{\thefootnote}{\arabic{footnote})}
\setcounter{footnote}{0}

\setcounter{equation}{0}
\section{{\mathversion{bold}$N=2$} supersymmetric KdV hierarchies: the third family}

\noindent
{\bf {\mathversion{bold}$N=2$} supersymmetry.}
We shall consider an $N=2$ superspace with space coordinate~$x$ and two
Grassmann coordinates $\theta$, $\bar\theta$. We shall use the
notation ${\underline x}$ for the triple of coordinates
$(x,\theta,\bar\theta)$. The supersymmetric covariant derivatives
are def\/ined by
\begin{equation}
\partial\equiv{\partial\over\partial x},\qquad
 D={\partial\over\partial\theta} +{1\over 2}\bar\theta\partial,\qquad
\bar D={\partial\over\partial\bar\theta} +{1\over 2}\theta\partial ,
\end{equation}
and satisfy the $N=2$ supersymmetry algebra
\begin{equation}
D^2=\bar D^2=0,\qquad \{ D,\bar D\}=\partial .
\label{n2alg}
\end{equation}
Beside ordinary superf\/ields $H({\underline x})$ depending
arbitrarily on Grassmann coordinates, one can also def\/ine chiral
superf\/ields $\varphi({\underline x})$ satisfying
$D\varphi =0$ and antichiral superf\/ields $\bar\varphi({\underline x})$
satisfying $\bar D\bar\varphi =0$.
We def\/ine the integration over the $N=2$ superspace to be
\begin{equation}
\int \mbox{d}^3{\underline x}\, H(x,\theta,\bar\theta)= \int \mbox{d}x\, \bar DDH(x,\theta,\bar\theta)
\vert_{\theta=\bar\theta=0}.
\end{equation}

Let us consider the algebra ${\cal D}$ of pseudo-dif\/ferential operators ${\cal L}$ of the form
\begin{equation}
{\cal L} = \sum_{k<M} u_{k}\partial^{k}
\label{1.1.01}
\end{equation}
with the usual product rule. The coef\/f\/icients functions $u_k$ are commuting $N=2$
superf\/ields. The highest power of $\partial$ with non zero coef\/f\/icient will be called the 
order of ${\cal L}$. We def\/ine as usual the residue of the pseudo-dif\/ferential operator 
${\cal L}$ by $\mbox{res}\, {\cal L}=u_{-1}$. The residue of a commutator is a total 
space derivative, $\mbox{res}\, [{\cal L}, {\cal L}']= (\partial \Omega)$. The trace 
of ${\cal L}$ is the integral over the superspace of the residue
\begin{equation}
\mbox{tr}\, {\cal L} = \int \mbox{d}^3 \underline{x} \; \mbox{res}\,{\cal L} , \qquad
\mbox{tr}\, [{\cal L},{\cal L}'] = 0.
\label{1.1.03}
\end{equation}
${\cal D}$ can be split into two associative subalgebras 
${\cal D}={\cal D}_{+} \oplus {\cal D}_{-}$, where ${\cal L}$ is in ${\cal D}_{+}$ if it is 
a dif\/ferential operator and ${\cal L}$ is in ${\cal D}_{-}$ if it is a strictly pseudo-dif\/ferential 
operator ($M=0$ in \ref{1.1.01}). We shall note
\begin{equation}
{\cal L} = {\cal L}_{+} + {\cal L}_{-} , \qquad {\cal L}_{\pm} \in {\cal D}_{\pm}.
\label{1.1.05}
\end{equation}
Moreover, ${\cal D}_{+}$ and ${\cal D}_{-}$ are isotropic subalgebras with respect to the trace.
As a consequence of these facts, the endomorphism $R$ of ${\cal D}$ def\/ined 
by $R({\cal L}) = {1 \over 2}({\cal L}_{+}-{\cal L}_{-})$ is a skew-symmetric classical 
$r$-matrix (the very same as in the bosonic case),
\be
\mbox{tr}\, (R({\cal L}){\cal L}'+{\cal L}R({\cal L}')) =0,
\label{1.1.07}
\ee
\be
R([R({\cal L}),{\cal L}']+[{\cal L},R({\cal L}')]) = [R({\cal L}),R({\cal L}')]
+{1 \over 4}[{\cal L},{\cal L}'].
\label{1.1.09}
\ee

\noindent
{\bf KP type equations.} 
We shall f\/irst write the evolution equations for an $N=2$
super\-sym\-metric KP type hierarchy. Let us consider a pseudo-dif\/ferential operator of the type
\begin{equation}
{\cal L} = \partial^{n-1} + \sum_{k=1}^{\infty} U_{k}\partial^{n-1-k}
\label{1.1.11}
\end{equation}
containing an inf\/inite set of bosonic $N=2$ superf\/ields. Following \cite{KS}, 
we associate with~${\cal L}$ the two pseudo-dif\/ferential 
operators\footnote{Our conventions dif\/fer from those of Krivonos and Sorin. In particular, 
chiralities are exchanged.}
\begin{equation}
L=\{ D ,{\cal L} \bar D {\cal L}^{-1} \} , \qquad
\bar L =\{ \bar D ,{\cal L}^{-1}  D {\cal L} \}.
\label{1.1.13}
\end{equation}
By def\/inition, $L$ and $\bar L$ are respectively chiral and anti-chiral operators, that is 
to say $[D,L]= [\bar D , \bar L]=0$. $L$ and $\bar L$ are pseudo-dif\/ferential operators 
in ${\cal D}$, of order one, which becomes clear when they are written in the form
\begin{equation}
L= \partial + [D {\cal L} [\bar D {\cal L}^{-1}]] , \qquad
\bar L = \partial + [ \bar D {\cal L}^{-1}  [D {\cal L}]],
\label{1.1.15}
\end{equation}
where the notation $[D{\cal L}]$ means that the odd derivative $D$ only acts on 
the coef\/f\/icient functions of the operator ${\cal L}$. $L$ and $\bar L$ are conjugate operators
\begin{equation}
L  = {\cal L} \bar L {\cal L}^{-1}
\label{1.1.17}
\end{equation}
and, as a consequence, the following basic commutation relation between powers 
of $L$ and $\bar L$ holds
\begin{equation}
L^{p} {\cal L} = {\cal L} \bar L^{p}.
\label{1.1.19}
\end{equation}
This suggests to def\/ine the following f\/lows on ${\cal L}$
\begin{equation}
{\partial \over \partial t_p}{\cal L} = R(L^p){\cal L}-{\cal L}R(\bar L^{p})
\label{1.1.21}
\end{equation}
which do not have the Lax form. Using equation (\ref{1.1.19}), these f\/lows can 
be put into the form
\begin{equation}
\partial_p{\cal L} = (L^p)_{-}{\cal L}-{\cal L}(\bar L^{p})_{-}.
\label{1.1.23}
\end{equation}
The right-hand side is  a $\Psi$DO of order $n-2$, which proves the consistency
of the f\/lows~(\ref{1.1.23}). We used the notation $\partial_p = \partial / \partial t_p$. 
A crucial point is that, although the f\/lows on ${\cal L}$ do not take the Lax form, 
the evolution equations of $L$ and $\bar L$ indeed do
\begin{equation}
\partial_{p} L = [R(L^p),L], \qquad
\partial_{p} \bar L = [R(\bar L^p),\bar L ].
\label{1.1.25}
\end{equation}
In order to show this, one can study the evolution equations of the operator
${\cal L}\bar D {\cal L}^{-1}$, which using  (\ref{1.1.21}) are given by
\begin{equation}
\partial_{p}{\cal L}\bar D {\cal L}^{-1}=[R(L^p),{\cal L}\bar D {\cal L}^{-1}]
+{\cal L}[\bar D, R(\bar L^{p})] {\cal L}^{-1}.
\end{equation}
The second term on the right-hand side vanishes because of the anti-chirality
of the ope\-ra\-tor $\bar L$. The evolution equations of $L$ are then obtained as
\begin{equation}
\partial_{p}L=\{ D,\partial_{p}{\cal L}\bar D {\cal L}^{-1}\}=
\{ D,[R(L^p),{\cal L}\bar D {\cal L}^{-1}]\}=[R(L^p),L],
\end{equation}
where the chirality of the operator $L$ has been used in the last equality.
Using the Lax equations (\ref{1.1.25}) for $L$ and $\bar L$ and the modif\/ied 
Yang-Baxter equation (\ref{1.1.09}) for $R$, one can show that the f\/lows (\ref{1.1.21}) commute
\be
\ba{l}
[\partial_p , \partial_q] {\cal L} =\left( R([R(L^q),L^p])-R([R(L^p),L^q])
+[R(L^p),R(L^q)]\right){\cal L} 
\vspace{2mm}\\
\qquad  -{\cal L}\left( R([R(\bar L^q),\bar L^p])-R([R(\bar L^p),\bar L^q])
+[R(\bar L^p),R(\bar L^q)]\right)= 0, 
\ea \label{1.1.27}
\ee
which suggests that this hierarchy is integrable.

To conclude this paragraph, we would like to mention a geometric formulation of the $N=2$
KP type f\/lows described before. Actually, one can def\/ine the following two sets of 
derivative operators
\be
\nabla_C = D , \qquad \bar\nabla_C = {\cal L}\bar D {\cal L}^{-1} ,
\qquad \partial_C = L, 
\ee
\be
\nabla_A = {\cal L}^{-1} \nabla_C {\cal L}={\cal L}^{-1} D {\cal L},
\qquad \bar\nabla_A = {\cal L}^{-1} \bar\nabla_C {\cal L}=\bar D,
\qquad \partial_A = {\cal L}^{-1} \partial_C  {\cal L}=\bar L
\ee
which both satisfy the $N=2$ supersymmetry algebra
\begin{equation}
\nabla_X^2 = \bar\nabla_X^2 =0, \qquad
 \partial_X = \{ \nabla_X , \bar\nabla_X \} \quad \mbox{for} \quad X = C, A.
\end{equation}
Hence, an easy computation shows that the KP type f\/lows (\ref{1.1.21}) imply a 
Lax type evolution equation for all these derivatives
\begin{equation}
\partial_p \nabla_X = [R(\partial_X^p),\nabla_X], \qquad
 \partial_p \bar\nabla_X = [R(\partial_X^p),\bar\nabla_X]\quad \mbox{for} \quad X = C, A.
\end{equation}

\noindent
{\bf A discrete symmetry.} The f\/lows of the KP type hierarchy are invariant 
under the following involutive transformation $T$:
\be
\ba{l}
 x_T=x,\qquad \theta_T=\bar\theta, \qquad \bar\theta_T=\theta,
\qquad t_{Tp}=(-)^{p+1} t_p,
\vspace{2mm}\\
 {\cal L}_{T}(x_T,\theta_T,\bar\theta_T)=(-)^{n-1} {\cal L}^{t}(x,\theta,\bar\theta),
\ea
\label{1.1.29}
\ee
where ${\cal L}^{t}$ is the adjoint operator, def\/ined as in the bosonic case \cite{Dickey}, 
and $n-1$ is the order of ${\cal L}$. Indeed the f\/lows (\ref{1.1.21}) are equivalent to
\begin{equation}
(-)^{p+1}\partial_p {\cal L}_{T} = R(L_T^p){\cal L}_{T}-{\cal L}_T R(\bar{L}_T^{p})
\label{1.1.31}
\end{equation}
where $L_T= \{ \bar D_T , {\cal L}_{T} D_T {\cal L}_{T}^{-1} \}$ and 
$\bar{L}_T = \{ D_T , {\cal L}_{T}^{-1} \bar D_T {\cal L}_{T} \}$. Using this result, certain 
discrete invariance for the nonlinear evolution equations \cite{KS} can be extracted 
directly from the operator ${\cal L}$. We shall see a simple example below.

\medskip

\noindent
{\bf Conserved quantities.} Another consequence of Lax equations (\ref{1.1.25}) 
is that they provide us with an inf\/inite set of conserved quantities for the f\/lows (\ref{1.1.21}). 
Standard arguments coming from the study of KP type hierarchies \cite{Dickey} tell us 
that conservation laws are associated with $\mbox{res}\, L^k$ and 
$\mbox{res}\, \bar L^k$ so that the quantities~\cite{bokris,KS}
\begin{equation}
H_{k}= {1\over k}\int \mbox{d} x \; \mbox{res}\; L^k \vert_{\theta , \bar\theta =0}= 
{1\over k} \int \mbox{d} x \; \mbox{res}\, \bar L^k \vert_{\theta , \bar\theta =0}
\label{1.2.01}
\end{equation}
are conserved. Notice here that the integration is over the space coordinate $x$ 
only, so that it is not clear a priori why these quantities are invariant under $N=2$
supersymmetry. This fact, as well as the second equality in (\ref{1.2.01}), are 
consequences of the basic relation~(\ref{1.1.19}) which yields
\begin{equation}
\mbox{res}\, \bar L^k =\mbox{res}\, L^k +\mbox{res}\, [{\cal L}^{-1}, L^{k}{\cal L} ]
\label{1.2.01a}
\end{equation}
so that, from the properties of the residue \cite{Dickey}, the quantities $\mbox{res}\, L^k$ 
and $\mbox{res} \, \bar L^k$ dif\/fer only by a space derivative, which we denote by
\begin{equation}
\mbox{res}\, [{\cal L}^{-1}, L^{k}{\cal L} ]=k\partial{\cal H}_k
\end{equation}
where ${\cal H}_k$ is an $N=2$ dif\/ferential polynomial in the coef\/f\/icients of ${\cal L}$. 
We may rewrite equation (\ref{1.2.01a}) as
\begin{equation}
\mbox{res}\, \bar L^k -k\bar DD{\cal H}_k=\mbox{res}\, L^k+kD\bar D{\cal H}_k.
\label{1.2.02}
\end{equation}
In this last equation, the left-hand side is an antichiral superf\/ield, whereas the 
right-hand side is a chiral superf\/ield. Then both sides must be equal to a constant, and we get
\begin{equation}
H_{k} = \int \mbox{d}^3\underline{x} \, {\cal H}_k
\label{1.2.03}
\end{equation}
This last expression of the conserved charges involves an integration on the whole $N=2$
superspace, which ensures that they are invariant under supersymmetry.

\medskip

\noindent
{\bf Hamiltonian structure.} 
We turn now to the problem of constructing a Poisson bracket for the f\/lows (\ref{1.1.21}). 
From the analysis of the third series of $N=2$
KdV hierarchies \cite{pluri,yung2}, we expect the existence of a single quadratic 
hamiltonian structure corresponding to the $N=2$
${\cal W}_{n}$ algebra. Since the f\/lows (\ref{1.1.21}) do not have the Lax form,
 we do not expect this quadratic bracket to be of the Adler-Gelfand-Dickey or the {\it abcd} 
type as is the case for the two f\/irst series of $N=2$
KdV hierarchies. Nevertheless, it is possible to use the quadratic hamiltonian structures 
found in \cite{DG1} in order to get one for the hierarchy we are dealing with. Let us 
give a brief account of some of the results in~\cite{DG1}.

We consider the associative algebra $\check{\cal C}$ of
pseudo-dif\/ferential operators $\check{L}$ preserving chirality of the form
\begin{equation}
\check{L}=D{\cal L}\bar D, \qquad {\cal L}= \sum_{i <M} u_{i}\partial^{i}\in {\cal D}.
\label{1.4.01}
\end{equation}
The coef\/f\/icient functions $u_i$ again are commuting $N=2$ superf\/ields\footnote{Although 
there is an obvious bijection between operators in ${\cal D}$ and 
operators in $\check{\cal C}$, the products in the two algebras dif\/fer.}.
We def\/ine the residue of the pseudo-dif\/ferential operator $\check{L}$ by
$\mbox{Res} \, \check{L}=\mbox{res}\, {\cal L}$. The residue of a commutator is a 
total derivative in $N=2$ superspace,
$\mbox{\rm Res}\, [\check{L},\check{L}']=D\bar\omega+\bar D\omega$. 
The trace of $\check{L}$ is the integral of the residue
\begin{equation}
\mbox{Tr}\, \check{L}=\int \mbox{d}^3{\underline x}\; \mbox{Res}\, \check{L}, \qquad
\mbox{Tr}\, [\check{L},\check{L}']=0.
\label{1.4.03}
\end{equation}
We def\/ine a classical $r$-matrix in
$\check{\cal C}$ by $\check{R}(\check{L})=DR({\cal L})\bar D$\footnote{The classical 
Yang-Baxter equation satisf\/ied by $\check{R}$ does not follow from this same equation for $R$.
It requires an independent proof.}. It is not skew symmetric, but satisf\/ies the equation
\begin{equation}
\mbox{Tr}\, (\check{R}(\check{L})\check{L}'+\check{L}\check{R}(\check{L}'))
=-\int \mbox{d}^3{\underline x} \; \mbox{Res}\, \check{L} \; \mbox{Res}\, \check{L}'.
\label{1.4.07}
\end{equation}
To this $r$-matrix correspond two a priori dif\/ferent quadratic Poisson brackets, 
which however are related by a Poisson map. In this article we shall use the f\/irst of these 
brackets. Let ${\check{X}}$  be some $\Psi$DO in $\check{\cal C}$ with coef\/f\/icients 
independent of the phase space f\/ields $\{u_i\}$, then def\/ine the linear functional
$l_{{\check{X}}}(\check{L}) = \mbox{Tr}\, (\check{L}{\check{X}})$. We also def\/ine projections 
$\Phi$ and $\bar\Phi$ on the chiral and antichiral parts of a general $N=2$ 
superf\/ield $H$ by\footnote{An explicit expression of the map $\Phi$ may be found 
in \cite{DG1}.}
\begin{equation}
H=\Phi(H)+\bar\Phi(H),\qquad D\Phi(H)=0,\qquad\bar D\bar\Phi(H)=0.
\end{equation}
The f\/irst quadratic bracket in \cite{DG1} then reads
\be
\ba{l}
\{ l_{{\check{X}}},l_{{\check{Y}}} \}_{(2)}^a (\check{L}) =\mbox{Tr}\, 
 \left( \check{L}{\check{X}}\check{R}(\check{L}{\check{Y}})-{\check{X}}
\check{L}\check{R}({\check{Y}}\check{L})\right.
\vspace{2mm}\\
\ds \phantom{\{ l_{{\check{X}}},l_{{\check{Y}}} \}_{(2)}^a (\check{L}) =}
 \left.+ \Phi(\mbox{Res}\, [\check{L},{\check{Y}}]) 
\check{L}{\check{X}}+{\check{X}}\check{L}
\bar\Phi(\mbox{Res}\, [\check{L},{\check{Y}}])\right),
\ea \label{1.4.09}
\ee

We now wish to rewrite this hamiltonian structure directly on ${\cal L}$. We consider linear
functionals on the phase space $l_{X}[{\cal L}]= \mbox{tr}\, ({\cal L}X)$
where $X$ is a pseudo-dif\/ferential operator in ${\cal D}$ independent of the phase space 
f\/ields $u_k$. We shall use the relation between linear functionals of ${\cal L}$ and $\check{L}$
\be
\check{L} = D {\cal L} \bar D , \qquad \check{X} = D \bar D \partial^{-1} X \partial^{-1}D \bar D,
\ee
\be
l_{X}[{\cal L}]= \mbox{tr}\, ({\cal L}X) = \mbox{Tr}\, (\check{L}\check{X}) = l_{\check{X}}[\check{L}].
\label{1.4.17}
\ee
One then obtains the hamiltonian structure
\be
\ba{l}
\{ l_{X}, l_{Y} \}({\cal L}) = \mbox{tr} \left( {\cal L}X\partial ({\cal L}Y)_{+}- 
{\cal L}X([D[\bar D {\cal L}]Y])_{+} +[ D {\cal L}[\bar D X]]({\cal L}Y)_{+}\right.
\vspace{2mm}\\
\phantom{\{ l_{X}, l_{Y} \}({\cal L}) =} -X{\cal L}\partial (Y{\cal L})_{+}   +X{\cal L}([\bar D [ D Y]{\cal L}])_{+}-
[\bar D X [ D {\cal L}]](Y{\cal L})_{+}
\vspace{2mm}\\
\phantom{\{ l_{X}, l_{Y} \}({\cal L}) =} \left. +  {\cal L}X\Phi(\mbox{res}\, [{\cal L},Y])+X{\cal L}\bar\Phi (\mbox{res}\,[{\cal L},Y])\right) .
\ea \label{1.4.19}
\ee
Let us insist here that the Poisson bracket (\ref{1.4.19}) is in fact identical
to (\ref{1.4.09}), which properties, and in particular the Jacobi identities, 
have been studied in \cite{DG1}.

We shall now show that this hamiltonian structure, together with the hamiltonian 
$H_p$, generates the f\/lows (\ref{1.1.21}). For any functional $F[{\cal L}]$, we def\/ine the
functional derivative ${\delta F \over \delta {\cal L}}\in{\cal D}$
as follows. Under a small variation $\delta {\cal L}$ of ${\cal L}$, the variation 
of the functional is
\begin{equation}
\delta F[{\cal L}] = \int \mbox{d}^3\underline{x} \; \mbox{res} \left( {\delta F \over \delta {\cal L}} 
\delta {\cal L}\right) = \mbox{tr}\left( {\delta F \over \delta {\cal L}} \delta {\cal L} \right) .
\label{1.4.21}
\end{equation}
Then we consider the hamiltonian $H_p$ def\/ined in equation (\ref{1.2.01}).
Using the fact that $L$ is chiral and $\bar L$ is anti-chiral, we obtain the functional derivative
\begin{equation}
{\delta H_p \over \delta {\cal L}} = \bar L^{p-1}{\cal L}^{-1} = {\cal L}^{-1}L^{p-1}.
\label{1.4.23}
\end{equation}
In order to compute the hamiltonian vector f\/ield associated with the functional $H_p$, 
we use the following identities
\be
\ba{l}
\ds \left[{\cal L},{\delta H_p \over \delta {\cal L}}\right]=L^{p-1}-\bar L^{p-1}
\vspace{2mm}\\
\ds \Longrightarrow \qquad \Phi\left(\mbox{res}\left[{\cal L},{\delta H_p \over \delta {\cal L}}
\right]\right) =\mbox{res}\, L^{p-1},\qquad
\bar\Phi\left(\mbox{res}\left[{\cal L},{\delta H_p \over \delta {\cal L}}\right]\right)
=-\mbox{res}\, \bar L^{p-1},
\vspace{2mm}\\
\ds \phantom{\Longrightarrow \qquad} (\partial L^{p-1})_+=\partial(L^{p-1})_++\mbox{res}\, L^{p-1}
\ea 
\ee
Using the expressions (\ref{1.1.15}), a fairly easy computation then leads to
\begin{equation}
\partial_p l_X[{\cal L}] = \{ l_X , H_p\}({\cal L})
= \mbox{tr} \left( X (R(L^p){\cal L}-{\cal L}R(\bar L^{p}))\right)
\label{1.4.25}
\end{equation}
which is the desired equation of motion. The Poisson bracket of two hamiltonians reads
\begin{equation}
\{ H_p , H_q\} =\mbox{tr} \left(L^{p-1}R(L^q)-\bar L^{p-1}R(\bar L^q)\right)=0.
\label{1.4.27}
\end{equation}
The last equality follows from the fact that the superspace integral of a chiral 
or antichiral superf\/ield vanishes. We thus proved the integrability of this hierarchy.

\medskip

\noindent
{\bf Reductions.} In order to f\/ind reductions of the KP type hierarchy, we need to 
f\/ind Poisson submanifolds of the KP type phase space. The Poisson submanifolds 
of the quadratic bracket (\ref{1.4.19}) correspond to those of the quadratic brackets 
$\{ , \}_{(2)}^{a}$ which were given in \cite{DG1}. In particular the constraint
\begin{equation}
{\cal L} = {\cal L}_{+}
\label{1.5.01}
\end{equation}
def\/ines a f\/irst type of Poisson submanifold. For an operator ${\cal L}$ of order $n-1$, 
the corresponding Poisson algebra is the $N=2$ ${\cal W}_n$ algebra and the hierarchy 
thus obtained is the third $N=2$ $n$-KdV hierarchy. The simplest example of such 
a hierarchy\footnote{Many examples of the f\/irst f\/lows of these hierarchies can be found 
in \cite{KS}.}  is provided by the choice
\begin{equation}
{\cal L} = \partial + J.
\label{1.5.03}
\end{equation}
The f\/irst non trivial f\/low is
\begin{equation}
\partial_2 J = \left( [D,\bar D]J-J^2\right)_x,
\label{1.5.05}
\end{equation}
where one recognizes the second f\/low of the so-called $N=2$ $a=4$ KdV
 hierarchy. Two alternative Lax operator for this hierarchy are already known
 \cite{mathieu,KSb,DG1}, but the reason why these three operators give the same 
conserved quantities is not clear to us. Notice that the transformation $T$ (\ref{1.1.29}) 
is given in this example by
\begin{equation}
J \rightarrow -J , \qquad D \leftrightarrow \bar D , \qquad t_2 \rightarrow - t_2 
\end{equation}
and is a symmetry of the f\/low (\ref{1.5.05}). The Poisson algebra for this hierarchy
can be directly worked out from the general hamiltonian structure (\ref{1.4.19}). With the 
choice $X=f\partial^{-1}$, $Y=g\partial^{-1}$ where $f$ and $g$ are superf\/ields 
independent on $J$, one obtains the following Poisson bracket between the two 
linear functionals $l_{X}({\cal L})= \int \mbox{d}^3{\underline x} \, Jf$, 
$l_{Y}({\cal L})= \int \mbox{d}^3{\underline x} \, Jg$
\begin{equation}
\biggl\{ \int Jf, \int Jg \biggr\}= \int \mbox{d}^3{\underline x} \; f \left( (Jg)_x -DJ \, 
\bar D g-\bar D J  \, Dg - [D, \bar D]g_x\right).
\end{equation}
One recognizes in this expression the classical $N=2$ superconformal algebra.

Another possible reduction is to take ${\cal L}$ of the form
\begin{equation}
{\cal L} = {\cal L}_{+} + \phi \partial^{-1} \bar\phi , \qquad D \phi = \bar D \bar \phi =0,
\end{equation}
where $\phi$ and $\bar\phi$ are Grassmann even or odd superf\/ields. In this case, 
the Poisson algebra is an extension of the $N=2$ ${\cal W}_n$ algebra for ${\cal L}$ of 
order $n-1$. This extension is non local if $\phi$ and $\bar\phi$ are Grassmann odd \cite{DG1}. 
The simplest example of such a hierarchy is provided by the choice
\begin{equation}
{\cal L} = 1 +\phi \partial^{-1} \bar\phi
\label{1.5.07}
\end{equation}
where $\phi$ and $\bar\phi$ are Grassmann odd. Krivonos and Sorin \cite{KS} 
noticed that the f\/irst non trivial f\/low belongs to the $N=2$ NLS hierarchy for which an
alternative Lax operator exists \cite{KST,DG1}. The relation between both operators is unclear. 
An other simple example is provided by the choice
\begin{equation}
{\cal L} = \partial+ J +\phi \partial^{-1} \bar\phi,
\label{1.5.09}
\end{equation}
where $\phi$ and $\bar\phi$ are Grassmann even. The Poisson algebra for this hierarchy is
the ``small'' $N=4$ superconformal algebra (SCA). The computation of the f\/irst f\/lows 
for this ope\-ra\-tor~\cite{KS} shows that this hierarchy is neither the ``small''
 $N=4$ KdV hierarchy, nor the ``quasi'' $N=4$ KdV \cite{DG1,DGI}. Hence, this is an 
other example of integrable hierarchy having the $N=4$ SCA as a hamiltonian structure 
but which respects only $N=2$ supersymmetry.

\medskip

\noindent
{\bf Bosonic limit.}
For completeness, we study the bosonic limit of the $N=2$ $n$-KdV hierarchy 
described before, which was already considered in \cite{KS}. One f\/inds, as it has been 
conjectured in \cite{bokris}, that this limit is actually the $(1,n)$ KdV hierarchy  
\cite{KS,aratyn,BX,Dickey2}.

From now on, we restrict to operators ${\cal K}$ in ${\cal D}$ satisfying the 
conditions $D {\cal K}\vert_{0}=\bar D {\cal K}\vert_{0}=0$, where the limit $\vert_{0}$ 
means that $\theta$ and $\bar\theta$ are set to zero. This def\/ines a 
subspace ${\cal D}_B$ of~${\cal D}$ which is closed under the usual product. 
To an operator ${\cal K}$ in ${\cal D}_B$ we can associate two ordinary bosonic operators 
in ${\cal D}_0$ by
\be
\pi_1({\cal K}) = {\cal K}_1 = {\cal K}\vert_{0},
 \label{1.6.01}
\ee
\be
\pi_2({\cal K}) = {\cal K}_2 = {\cal K}\vert_{0} \partial + [\bar D D {\cal K}]\vert_{0}.
 \label{1.6.02}
\ee
Remark that, if $k$ is the order of ${\cal K}$, then the respective orders of ${\cal K}_1$ 
and ${\cal K}_2$ are $k$ and $k+1$. It is easily checked that $\pi_1$ is a morphism 
from ${\cal D}_B$ to ${\cal D}_0$, that is to say $({\cal K}{\cal K}')_1={\cal K}_1{\cal K}_1'$, 
and that the following property holds
\begin{equation}
({\cal K}_+)_1 = ({\cal K}_1)_+  , \qquad  ({\cal K}_-)_1 = ({\cal K}_1)_-.\label{1.6.03}
\end{equation}
It may be checked, from their def\/inition (\ref{1.1.13}), that the limit of the 
supersymmetric operators $L$ and $\bar L$ is given by
\begin{equation}
L_1 = {\cal L}_2 {\cal L}_1^{-1} , \qquad \bar L_1 = {\cal L}_1^{-1} {\cal L}_2. \label{1.6.04}
\end{equation}
Using these properties, it may be shown that the f\/lows (\ref{1.1.21}) have the following limit
\be
\partial_p {\cal L}_1 = R(L_1^p){\cal L}_1-{\cal L}_1 R(\bar L_1^p),
 \label{1.6.05}
\ee
\be
\partial_p {\cal L}_2 = R(L_1^p){\cal L}_2-{\cal L}_2 R(\bar L_1^p)
\label{1.6.06}
\ee
which are precisely the def\/ining f\/lows of the $(1,n)$ hierarchy, once one has restricted 
${\cal L}$ to be a dif\/ferential operator of order $n-1$. The hamiltonian structure for this 
system can be recovered from the supersymmetric Poisson bracket (\ref{1.4.19}) following 
the lines of \cite{DG1}.

\medskip

\noindent
{\bf Wronskian solutions.}
Our f\/irst goal in this paragraph will be to construct solutions of the
nonlinear equations (\ref{1.1.23}) in terms of a set of functions satisfying linear equations. 
First remark that the f\/low equations (\ref{1.1.19}) and
 (\ref{1.1.21}) for $L$ and $\bar L$ are just the standard KP f\/lows. Simply, the coef\/f\/icient 
functions in $L$ and $\bar L$ are not ordinary functions, but rather constrained superf\/ields. 
Then, it is reasonable to introduce a set of $P$ chiral superf\/ields $Y_i$ and a set of $Q$ 
antichiral superf\/ields $\bar Y_i$ satisfying
\be
\ba{l}
DY_i=0,\qquad \partial_k Y_i=\partial^k Y_i,\qquad i=1,\ldots, P,
\vspace{2mm}\\
\bar D\bar Y_i=0,\qquad \partial_k \bar Y_i=\partial^k \bar Y_i,\qquad i=1,\ldots, Q.
\ea \label{KPcond}
\ee
We require the functions $Y_i$ (respectively the functions $\bar Y_i$) 
to be independent, that is to say that the Wronskians ${\cal W}(Y_1,\ldots, Y_P)$
and ${\cal W}(\bar Y_1,\ldots,\bar Y_P)$ do not vanish. Next, we introduce the 
dif\/ferential operators
\be
\ba{l}
\ds \Phi=\frac{1}{{\cal W}(Y_1,\ldots, Y_P)}\left|\begin{array}{cccc}
Y_1 & \cdots & Y_P & 1 \\
Y_1^{(1)} & \cdots & Y_P^{(1)} & \partial \\
\vdots & \ddots & \vdots & \vdots \\
Y_1^{(P)} & \cdots & Y_P^{(P)} & \partial^P \end{array}\right|,\qquad [D,\Phi]=0,
\vspace{4mm}\\
\ds \bar \Phi=\frac{1}{{\cal W}(\bar Y_1,\ldots, \bar Y_Q)}\left|\begin{array}{cccc}
\bar Y_1 & \cdots & \bar Y_Q & 1 \\
\bar Y_1^{(1)} & \cdots & \bar Y_Q^{(1)} & \partial \\
\vdots & \ddots & \vdots & \vdots \\
\bar Y_1^{(Q)} & \cdots & \bar Y_Q^{(Q)} & \partial^Q \end{array}\right|,
\qquad [\bar D,\bar \Phi]=0,
\ea
\ee
for which we derive in the usual way \cite{Dickey} the f\/low equations
\begin{equation}
\partial_k \Phi=(\Phi\partial^k\Phi^{-1})_-\Phi,\qquad
\partial_k \bar \Phi=(\bar \Phi\partial^k\bar \Phi^{-1})_- \bar \Phi.
\label{eqq}
\end{equation}
Then we consider the dressed operator
\begin{equation}
{\cal L}=\Phi\partial^p\bar \Phi^{-1},\label{sol}
\end{equation}
where the order of ${\cal L}$ is $n$ and $p=n+Q-P$ is a positive integer,
and derive the expressions
\begin{equation}
L=\{ D,{\cal L}\bar D{\cal L}^{-1}\}= \Phi\partial\Phi^{-1},\qquad
\bar L=\{\bar  D, {\cal L}^{-1}D{\cal L}\}=
\bar\Phi\partial\bar\Phi^{-1}.
\label{habil}\end{equation}
We conclude, using (\ref{eqq}), that the operator ${\cal L}$ def\/ined in equation (\ref{sol}),
satisfy the KP type f\/low equations (\ref{1.1.23}).

We now wish to restrict to KdV type f\/lows. In other words we wish to f\/ind conditions 
which ensure that ${\cal L}$ is a dif\/ferential operator. In order to do this, we borrow 
from \cite{oevel} the formula
\begin{equation}
\bar \Phi^{-1}=\sum_{i=1}^Q\bar Y_i\partial^{-1} \bar Z_i,\qquad
\bar Z_i=(-)^{Q-i}\frac{{\cal W}(\bar Y_1,\ldots,\bar Y_{i-1},\bar Y_{i+1},\ldots, 
\bar Y_Q)}{{\cal W}(\bar Y_1,\ldots, \bar Y_Q)}
\label{oevstr}\end{equation}
from which we deduce that the pseudo-dif\/ferential part of $\cal L$ in (\ref{sol}) reads
\begin{equation}
{\cal L}_-=\sum_{i=1}^Q[\Phi \bar Y_i^{(p)}]\partial^{-1} \bar Z_i.
\end{equation}
It is then clear that suf\/f\/icient conditions for $\cal L$ to be a dif\/ferential operator are
\begin{equation}
[\Phi \bar Y_i^{(p)}]={\cal W}(Y_1,\ldots,Y_P,\bar Y_i^{(p)})=0, \qquad i=1,\ldots,Q.
\label{KdVcond}
\end{equation}
It is to be noted that these equations mix chiral ($Y_i$) and antichiral ($\bar Y_i$) superf\/ields.

We shall now f\/ind a $\tau$-function for the $N=2$ KP type hierarchy, that is to say a 
generating function for the conserved quantities \cite{Dickey}. The starting point is the 
relation~(\ref{1.2.02}) between the hamiltonian density ${\cal H}_k$ and the residues 
of powers of $L$ and $\bar L$ which can be written as
\begin{equation}
k \partial {\cal H}_k = \mbox{res}\,  \bar L^k -\mbox{res}\, L^k.
\label{DHres}
\end{equation}
Our goal is then to obtain expressions for $\mbox{res}\, L^k$ and $\mbox{res}\, \bar L^k$
from the dressing relation
\begin{equation}
L^k= \Phi \partial^k \Phi^{-1},\qquad
\bar L^k= \bar\Phi\partial^k\bar\Phi^{-1}.
\end{equation}
For this we apply the formula (\ref{oevstr}) to $\Phi$ and $\bar \Phi$ and deduce
\begin{equation}
\mbox{res}\, L^k = \sum_{i=1}^{P} (-)^{P-i} [ \Phi  Y_i^{(k)} ] { {\cal W}[Y_1, \ldots , Y_{i-1},Y_{i+1},
\ldots , Y_P] \over {\cal W}[Y_1, \ldots , Y_P]}
\end{equation}
together with a similar result for $\mbox{res}\, \bar L^k$. We then make use of the 
Wronskian identity
\begin{equation}
{ {\cal W}[f_1, \ldots , f_{P+1}]{\cal W}[f_1, \ldots , f_{P-1}]\over {\cal W}[f_1, \ldots , f_{P}]^2}
= \left( { {\cal W}[f_1, \ldots , f_{P-1},f_{P+1}]\over {\cal W}[f_1, \ldots , f_{P-1},f_{P}]} \right)_{x}
\end{equation}
and obtain
\begin{equation}
\mbox{res} \, L^k = \sum_{i=1}^{P} \left( { {\cal W}[Y_1, \ldots ,Y_{i-1},Y_i^{(k)},Y_{i+1}, \ldots 
Y_P]\over {\cal W}[Y_1, \ldots , Y_{P}]} \right)_{x}.
\end{equation}
The result is then obtained by using the condition $\partial_k Y_i=\partial^k Y_i$ and reads
\begin{equation}
\mbox{res} L^k =  \partial \partial_k \ln {\cal W}[Y_1, \ldots , Y_{P}], \qquad 
\mbox{res}\, \bar L^k =  \partial \partial_k \ln {\cal W}[\bar Y_1, \ldots , \bar Y_{Q}].
\end{equation}
Finally, we get the following expression for the $\tau$-function
\begin{equation}
\tau (t,\theta , \bar\theta ) = { {\cal W}[\bar Y_1, \ldots , \bar Y_{Q}] \over {\cal W}[Y_1, \ldots , 
Y_{P}] } ,
\label{taufonc}
\end{equation}
which generates the hamiltonian densities according to
\begin{equation}
{\cal H}_k = {1 \over k} \partial_k \ln \tau  .
\label{genham}
\end{equation}

In the following, we shall give an example of a soliton solution constructed 
as described before for the $N=2$ $a=4$ KdV hierarchy. This case, which has been 
presented before, corresponds to the choice of the dif\/ferential operator ${\cal L}=\partial +J$. 
Here and further we use the notations
\be
\ds \zeta(t,\theta ,\bar\theta ; z, \mu ) = z\left(t_1-{1 \over 2}\theta\bar\theta\right) 
+ \sum_{k=2}^{\infty} z^k t_k + \mu \bar\theta, \qquad D\zeta=0,
\ee
\be
\ds \bar\zeta(t,\theta ,\bar\theta ; z, \mu ) = z\left(t_1+{1 \over 2}\theta\bar\theta\right) 
+ \sum_{k=2}^{\infty} z^k t_k + \mu \theta, \qquad \bar D \bar\zeta=0,
\ee
where $z$ is a real number, $\mu$ an odd Grassmann variable and $t_k$ are the times 
of the KdV hierarchy with the space variable $x$ identif\/ied with $t_1$.

We shall choose the number of chiral functions $Y$ to be $N=2$ and the number of 
antichiral functions $\bar Y$ to be $Q=1$ so that the integer $p=n+Q-P$ has value zero. 
We then def\/ine the two chiral functions to be
\begin{equation}
Y_1 = e^{\zeta(t,\theta ,\bar\theta ; z_1, \mu_1 )} + e^{\zeta(t,\theta ,\bar\theta ; z_2, \mu_2 )} , 
\qquad Y_2 =(Y_1)_x
\end{equation}
and the antichiral one to be
\begin{equation}
\bar Y_1 = e^{\bar\zeta(t,\theta ,\bar\theta ; z_1, \bar\mu_1 )} + 
e^{\bar\zeta(t,\theta ,\bar\theta ; z_2, \bar\mu_2 )}.
\end{equation}
A computation shows that this set of functions satisf\/ies the conditions (\ref{KPcond}) 
and (\ref{KdVcond}). The components of the KdV superf\/ield
\begin{equation}
J= J_0 +\theta J_{\theta}+\bar\theta J_{\bar\theta}+\theta\bar\theta J_{\theta\bar\theta}
\end{equation}
then read
\be
\ba{l}
\ds J_0 = \sqrt{z_1 z_2} { \mbox{ch}\,  ( \eta_1 + \ln z_1 -\eta_2 - \ln z_2 )/2  \over \mbox{ch}\,
 (\eta_1  -\eta_2 )/2 } , \qquad J_{\theta\bar\theta}= { (z_1-z_2)^2 \over 8 \, \mbox{ch}^2 
(\eta_1  -\eta_2 )/2 }  , 
\vspace{3mm}\\
\ds J_{\theta} = { (z_2-z_1)(\bar\mu_1-\bar\mu_2) \over 4 \, \mbox{ch}^2 (\eta_1  -\eta_2 )/2  }  , 
\qquad   J_{\bar\theta}=0.
\ea
\label{soliton}
\ee
In this expressions, we used the short-hand notation
\begin{equation}
\eta_k=\zeta(t,0,0 ; z_k, 0 )=\bar\zeta(t,0,0 ; z_k, 0 )
\end{equation}
for $k=1,2$. It is to be noted that, since only the dif\/ference $\eta_1-\eta_2$ appears in
the above equation, this solution represents a one-soliton for the $N=2$ $a=4$ KdV hierarchy. 
It may also be noted that if one sets $\bar\mu_1$ and $\bar\mu_2$ to zero in 
equation (\ref{soliton}), one obtains a soliton solution of the $(1,2)$ KdV hierarchy. 
Finally, the $\tau$-function associated with this solution~(\ref{taufonc}) reads
\begin{equation}
\tau (t,\theta , \bar\theta) = { e^{\bar\zeta(t,\theta ,\bar\theta ; z_1, \bar\mu_1 )} + 
e^{\bar\zeta(t,\theta ,\bar\theta ; z_2, \bar\mu_2 )} \over ( z_1 -z_2)^2 }
e^{- \zeta(t,\theta ,\bar\theta ; z_1, \mu_1 )-\zeta(t,\theta ,\bar\theta ; z_2, \mu_2 ) }
\end{equation}
and inspection of equation (\ref{genham}) shows that the hamiltonians are
\begin{equation}
H_k = { | z_1^k-z_2^k | \over k }.
\end{equation}

\subsection*{Acknowledgements}

This work is supported in part by CNRS, by the Emergence program of the 
Rh\^one-Alpes region (France) and by the EC-TMR contract FMRX-CT96-0012.

\label{delduc-lp}


\begin{thebibliography}{99}

\footnotesize


\bibitem{mathieu} Labelle P. and Mathieu P., {\it J. Math. Phys.}, 1991, V.32, 923.

\bibitem{pluri} Yung C.M., {\it Phys. Lett. B}, 1993, V.309, 75.\\
Bellucci S., Ivanov E., Krivonos S. and Pichugin A., {\it Phys. Lett. B}, 1993, V.312, 463.\\
Popowicz Z., {\it Phys. Lett. B}, 1993, V.319, 478.

\bibitem{yung2} Yung C.M. and Warner R.C., {\it J. Math. Phys.}, 1993, V.34, 4050.


\bibitem{DG1} Delduc F. and Gallot L., {\it Comm. Math. Phys.}, 1997, V.190, 395.


\bibitem{IK} Inami T. and Kanno H., {\it Int. J. Mod. Phys. A}, 1992, V.7, 419.

\bibitem{DG2} Delduc F. and Gallot L., Supersymmetric Drinfeld-Sokolov Reduction, 
solv-int/9802013.


\bibitem{bokris} Bonora L., Krivonos S. and Sorin A., {\it Nucl. Phys. B}, 1996, V.477, 835.

\bibitem{aratyn} Aratyn H., Nissimov E. and Pacheva S., {\it Phys. Lett. B}, 1993, V.314, 41.

\bibitem{BX} Bonora L. and Xiong C.S., {\it J. Math. Phys.}, 1994, V.35, 5781.\\
Bonora L. and Xiong C.S., {\it Comm. Math. Phys.}, 1996, V.175, 177.

\bibitem{Dickey2} Dickey L.A., {\it Lett. Math. Phys.}, 1995, V.35, 229.\\
Dickey L.A., On the Constrained KP Hierarchy II. An Additionnal Remark, hep-th/9411157.

\bibitem{KS} Krivonos S. and Sorin A.,  Extended $N=2$ Supersymmetric 
Matrix $(1,s)$-KdV Hierarchies, solv-int/9712002.

\bibitem{Dickey} Dickey L.A., Soliton Equations and Hamiltonian Systems, 
Adv. Ser. Math. Phys., Vol.12, World Scientif\/ic, 1991.

\bibitem{KSb} Krivonos S. and Sorin A., {\it Phys. Lett. B}, 1995, V.357, 94.

\bibitem{KST} Krivonos S., Sorin A. and Toppan F., {\it Phys. Lett. A}, 1995, V.206, 146.

\bibitem{DGI} Delduc F., Gallot L. and Ivanov E., {\it Phys. Lett. B}, 1997, V.396, 122.

\bibitem{oevel} Oevel W. and Strampp W., {\it J. Math. Phys.}, 1996, V.37, 6213.


\end{thebibliography}
\end{document}